\documentclass{article}

\usepackage[preprint]{neurips_2025}
\usepackage{graphicx}
\usepackage{amsmath}
\usepackage{makecell}

\usepackage[utf8]{inputenc} 
\usepackage[T1]{fontenc}    
\usepackage{hyperref}       
\usepackage{url}            
\usepackage{booktabs}       
\usepackage{amsfonts}       
\usepackage{nicefrac}       
\usepackage{microtype}      
\usepackage{xcolor}         

\title{Re-evaluating Position and Velocity Decoding for Hand Pose Estimation with Surface Electromyography}
\author{Nima Hadidi, Johannes Lee, Ebrahim Feghhi, Michael Yuan, Jonathan C. Kao}

\begin{document}
%
\maketitle

\begin{abstract}
Recent progress in real-time hand pose estimation from surface electromyography (sEMG) has been driven by the emg2pose benchmark, whose original baseline study concluded that velocity decoding outperforms position decoding in both reconstruction accuracy and trajectory smoothness. We revisit that conclusion under the original causal evaluation protocol. Using the same core architecture but a more stable training recipe, we show that position decoding models were previously underestimated because they are highly sensitive to a previously unswept decoder output scalar and can otherwise collapse into low movement solutions. Once this scalar is tuned, position decoding outperforms velocity decoding on the Tracking task across all three emg2pose generalization conditions, consistent with greater robustness to error accumulation. On the Regression task, the gap between position and velocity decoding is much smaller; instead, the largest gains come from multi-task training with Tracking, suggesting that the Regression objective alone does not sufficiently constrain the learned dynamics. Although position decoding models exhibit greater local jitter, a causal speed-adaptive filter preserves their accuracy advantage while yielding a more favorable smoothness-accuracy tradeoff than velocity decoding. Altogether, our results revise the original emg2pose modeling conclusions and establish a new state of the art among published streaming-compatible models on this benchmark.
\end{abstract}

\section{Introduction}

Real-time hand pose estimation from surface electromyography (sEMG) is a promising capability for both human-computer interaction and prosthetic control. Recent progress in this area has been enabled by the release of the emg2pose benchmark \citep{salter-2024}, which provides hundreds of hours of paired sEMG-pose recordings from hundreds of users and is the first resource of this scale for continuous hand pose estimation.

emg2pose evaluates generalization across held-out users, held-out prompted movement sequences (``stages''), and held-out stages on held-out users, with the last setting being the most relevant for practical deployment. It defines two related tasks. In \textit{Tracking}, models are given the ground-truth initial pose for each 5-second window and must predict the subsequent trajectory. In \textit{Regression}, models must predict the full trajectory without access to the ground-truth initial pose. These tasks capture distinct use cases: Tracking is well matched to multimodal systems in which sEMG complements another source of pose information such as computer vision, while Regression is more directly relevant to purely sEMG-based control.

A central conclusion of the original emg2pose baseline study was that \textit{velocity decoding} is preferable to \textit{position decoding} for real-time pose prediction. In Salter et al.'s vemg2pose model, the decoder predicts pose increments that are integrated over time, rather than directly predicting joint angles at each step. \citet{salter-2024} reported that this output parameterization improved both Tracking accuracy and trajectory smoothness relative to direct position decoding. If correct, this would be an important modeling lesson for streaming sEMG decoders.

However, the reported Tracking advantage of velocity decoding is not obviously intuitive. Because velocity decoding updates the current estimate by integrating incremental predictions over time, it is inherently exposed to error accumulation: each new prediction must implicitly compensate for past mistakes by adjusting the model's output distribution. Direct position decoding, by contrast, can in principle map the current sEMG context directly to the current pose without having to adjust the output distribution to correct for previous errors. The reported smoothness advantage of velocity decoding is easier to understand, since temporal integration acts as an implicit low-pass filter. But this does not by itself establish velocity decoding as the preferable modeling choice: if position decoding is more accurate, then smoothness may be better handled directly with lightweight causal post-processing rather than by accepting a less accurate output parameterization.

In this work, we revisit the position-versus-velocity question under the original causal emg2pose evaluation protocol. Using the same core architecture as \citet{salter-2024} but a more stable training recipe, we show that LSTM-based position decoding models are highly sensitive to a previously unswept decoder output scalar and can otherwise collapse into low-movement solutions. Once this scalar is tuned, position decoding outperforms velocity decoding on the Tracking task across all three emg2pose generalization conditions. On the Regression task, the difference between position and velocity decoding is much smaller; instead, the dominant effect is that augmenting Regression training with Tracking substantially improves performance for both output parameterizations, suggesting that the Regression objective alone does not sufficiently constrain the learned dynamics. Finally, although position decoding produces more local jitter than velocity decoding, we show that a simple adaptive causal low-pass filter preserves its accuracy advantage while yielding a more favorable smoothness-accuracy tradeoff. Taken together, these results revise the benchmark's original modeling conclusions and establish a new state-of-the-art among published streaming-compatible models on emg2pose.

\section{Related work}
\label{sec:format}

Since \citet{salter-2024} introduced the emg2pose dataset, follow-up work has largely moved away from the original real-time continuous pose-estimation benchmark: \citet{verma2025emg2tendon} extends the problem to tendon-control prediction and introduces a bidirectional diffusion-based model; \citet{lovecchiovq} studies tokenized pose decoding on a subset with custom splits and bidirectional decoding; \citet{cui2025cpep} and \citet{cuiembridge} use emg2pose for pose-informed EMG representation learning and evaluate gesture classification; \citet{liupose} uses emg2pose for multimodal pretraining in a different downstream EMG task; and \citet{chen2025physiowave}, \citet{fasulo2025tinymyo}, \citet{barmpas2025neurorvq}, and \citet{wang2025new} use emg2pose primarily as part of broader pretraining, tokenization/reconstruction, or generative pipelines rather than the original Tracking/Regression evaluation. Thus, despite substantial follow-up activity, we are not aware of any prior emg2pose follow-up that presents a causal, streaming-compatible model, and few works evaluate the original pose benchmark in a way that is directly comparable to the published protocol.

\section{Methods}
\subsection{Task and Notation}
Given a window of $C{=}16$-channel EMG sampled at $2\,\mathrm{kHz}$, $\mathbf{x}\in\mathbb{R}^{C\times T}$, we predict hand pose as $J{=}20$ joint angles $\mathbf{y}\in\mathbb{R}^{J\times T}$. For training/evaluation on the tracking task, we provide an initial pose $\mathbf{y}_0$ (taken from the first supervised time index of the window) and ignore frames marked as IK failures via a boolean mask.

\subsection{Core Architecture}
Our core architecture is identical to that of Salter et al. 2024.

\textbf{Encoder.} We use a causal 1D convolutional + TDS encoder. Two strided temporal conv blocks map EMG to 256 channels (kernel/stride $11/5$ then $5/2$). This is followed by two Time-Depth Separable (TDS) stages (each preceded by a subsampling conv with kernel/stride $17/4$ and $9/2$, respectively). The final stage outputs a $64$-D feature sequence
$\mathbf{f}_{1:K}\in\mathbb{R}^{64\times K}$ at a reduced temporal rate of $25\,\mathrm{Hz}$

\textbf{State-conditioned decoder.} We decode autoregressively at a rollout rate of $50\,\mathrm{Hz}$ by linearly interpolating encoder features to that rate. At each step $t$, the decoder input concatenates features and the previous pose estimate:
\[
\mathbf{z}_t = [\mathbf{f}_t;\,\hat{\mathbf{y}}_{t-1}] \in \mathbb{R}^{(64+20)}.
\]
The decoder is a 2-layer LSTM (hidden size 512) followed by a small output MLP. Its output is explicitly scaled by a fixed scalar $s$:
\[
\mathbf{o}_t = s\cdot g(\mathbf{h}_t),
\]
where $g(\cdot)$ is the decoder’s output projection (with a LeakyReLU before the final linear layer). Decoder predictions at $50\,\mathrm{Hz}$ are then linearly upsampled back to the EMG/joint-angle sample rate for supervision and metrics.

\subsection{Position vs. Velocity Variants}
Both variants share the same encoder and the same state-conditioned LSTM decoder; they differ only in how the decoder output $\mathbf{o}_t$ is interpreted:

\textbf{(i) Position decoding model.} The decoder directly outputs absolute joint angles:
\[
\hat{\mathbf{y}}_t = \mathbf{o}_t.
\]
Here, the output scalar $s$ globally scales the \emph{pose} magnitude produced by the decoder head.

\textbf{(ii) Velocity decoding model.} The decoder outputs an incremental update (discrete-time velocity), which is integrated over time:
\[
\Delta\hat{\mathbf{y}}_t = \mathbf{o}_t,\qquad
\hat{\mathbf{y}}_t = \hat{\mathbf{y}}_{t-1} + \Delta\hat{\mathbf{y}}_t,\quad \hat{\mathbf{y}}_0=\mathbf{y}_0.
\]
In this case, the same scalar $s$ scales the \emph{increment} size at every step (and therefore controls the effective step-to-step integration magnitude).

\subsection{Tasks and Single- vs. Multi-task Training}
We train and evaluate the same architecture under two task formulations that differ only in the supervision/setup of the autoregressive rollout.

\textbf{Tracking task.} The model is given the ground-truth initial pose at the first supervised frame, $\mathbf{y}_0$, and must continue the trajectory from that state. This is the standard state-initialized setting:
\[
\hat{\mathbf{y}}_0 = \mathbf{y}_0.
\]

\textbf{Regression task.} The model is not given the ground-truth initial pose, and must infer the entire pose trajectory from EMG alone. In this case, the first pose fed to the decoder is a learned vector
\[
\hat{\mathbf{y}}_0 = \mathbf{p}_{\mathrm{init}} \in \mathbb{R}^{J},
\]
shared across examples. This is a slight departure from Salter et al., who use a fixed zero vector. We found it preferable to make this initialization learnable because the all-zero joint-angle vector corresponds to a valid in-domain hand pose (approximately a flat hand), and thus is not a neutral ``no-information'' state.

For \emph{velocity decoding} models on the \textbf{Regression} task, we use a hybrid rollout (following \citet{salter-2024}) for the first $\tau_{\mathrm{pos}}=250\,\mathrm{ms}$ ($500$ samples at $2\,\mathrm{kHz}$): a separate position decoding head outputs absolute pose directly during this initial interval, after which the model switches to the velocity head and standard autoregressive integration. That is,
\[
\hat{\mathbf{y}}_t =
\begin{cases}
g_{\mathrm{pos}}(\mathbf{h}_t), & t \le \tau_{\mathrm{pos}},\\[4pt]
\hat{\mathbf{y}}_{t-1} + g_{\mathrm{vel}}(\mathbf{h}_t), & t > \tau_{\mathrm{pos}}.
\end{cases}
\]
In multitask models, this position-head warm start is used only for the \textbf{Regression} branch; the \textbf{Tracking} branch remains standard velocity decoding initialized from the provided $\mathbf{y}_0$.

\textbf{Single-task training.} In the single-task setting, we optimize the model for only one task (Tracking or Regression), using the corresponding rollout configuration throughout training.

\textbf{Multi-task training.} In the multi-task setting, the Tracking and Regression tasks share the same encoder and decoder parameters, but use task-specific rollout settings as above. Training minimizes a weighted sum of the two task losses,
\[
\mathcal{L}_{\mathrm{multi}} =
w_{\mathrm{track}}\,\mathcal{L}_{\mathrm{track}}
+
w_{\mathrm{reg}}\,\mathcal{L}_{\mathrm{reg}}.
\]
For the final multi-task models reported here, we use $w_{\mathrm{track}}=0.875$ and $w_{\mathrm{reg}}=0.125$. We also swept $w_{\mathrm{track}}=0.75$ and $w_{\mathrm{reg}}=0.25$ for all multi-task models, but generally found that this performed worse for all of our models.

\subsection{Optimization and Loss}
We train with AdamW (using the PyTorch default weight decay value of $10^{-2}$; weight decay is not applied to LayerNorm parameters) with gradient clipping (1.0). We adopt the learning rate schedule of \citet{sivakumar-2024}: linear warmup (10 epochs) from  $10^{-8}$ to $10^{-3}$, followed by cosine decay to $10^{-6}$ (140 epochs; 150 epochs total). The training loss is a weighted sum of joint-angle MAE and a forward-kinematics fingertip distance term:
\[
\mathcal{L} = \|\hat{\mathbf{y}}-\mathbf{y}\|_1
\;+\; \lambda_{\mathrm{tip}}\;\frac{1}{|\mathcal{T}|}\sum_{t\in\mathcal{T}}
\big\|\mathrm{FK}_{\mathrm{tip}}(\hat{\mathbf{y}}_t)-\mathrm{FK}_{\mathrm{tip}}(\mathbf{y}_t)\big\|_2,
\quad \lambda_{\mathrm{tip}}=0.01,
\]
(where $\mathcal{T}$ denotes a temporally downsampled set of valid frames to reduce compute). We apply a rotation augmentation to EMG during training.

\subsection{Causal speed-adaptive filtering}
\label{filter_method}
To control trajectory smoothness at inference time, we apply a causal,
speed-adaptive low-pass filter independently to each joint angle.
Given predictions $x_t$, the filtered output is

\begin{equation}
\hat{x}_t = \alpha_t x_t + (1-\alpha_t)\hat{x}_{t-1},
\end{equation}

where the smoothing factor depends on the instantaneous step magnitude,

\begin{equation}
v_t = \frac{|x_t - \hat{x}_{t-1}|}{T_e}, \qquad
\alpha_t = \frac{2\pi \beta v_t T_e}{1 + 2\pi \beta v_t T_e}.
\end{equation}

Here $T_e$ is the sampling period and $\beta$ is a single parameter that
controls the smoothness–accuracy trade-off. Small motions are strongly
suppressed ($\alpha_t \approx 0$), while large motions pass with minimal
attenuation ($\alpha_t \to 1$).

\section{Results}
\subsection{Position decoding requires appropriate output scaling to avoid low-movement solutions}

\begin{figure}[t!]
  \centering
  \includegraphics[height=8cm]{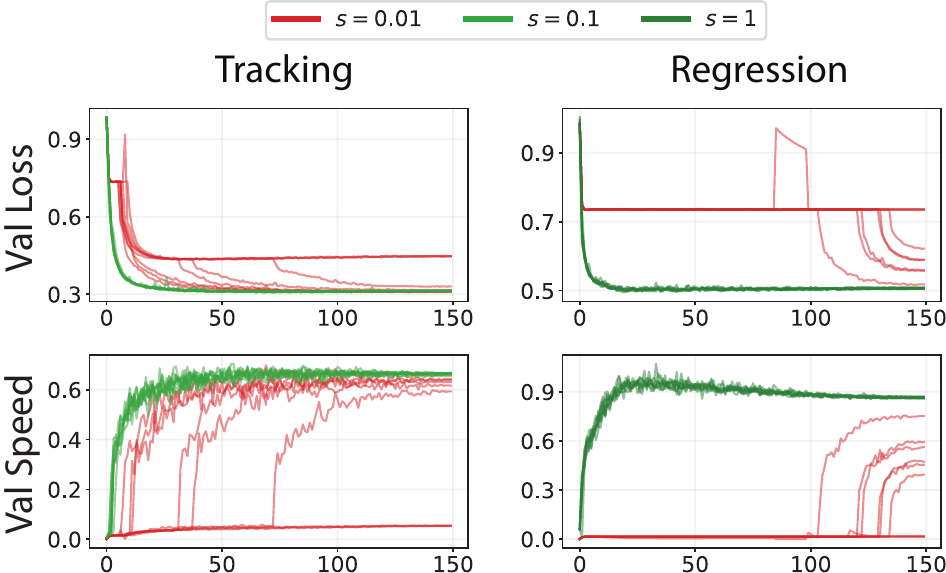}
  \caption{\textbf{Validation trajectories reveal that inappropriate output scaling can trap position decoding models in low-movement solutions.}
  Validation loss (top) and mean predicted speed (bottom) across epochs for position decoding models trained on Tracking (left) and Regression (right) with different decoder output scalars $s$. Thin lines show individual seeds. With the default scalar used by \citet{salter-2024} ($s=0.01$; red), training is often unstable and frequently converges to low-movement solutions, reflected in persistently low predicted speed. Increasing the scalar yields stable convergence for Tracking at $s=0.1$ and for Regression at $s=1$, indicating that position decoding performance is highly sensitive to output scaling.}
  \label{fig:fig1}
\end{figure}

Previously, \citet{salter-2024} reported that velocity decoding outperformed position decoding on the Tracking task. However, several details of their experiments suggest that position decoding models were trained under an unstable optimization regime. First, their hyperparameter sweep included a choice between an MLP and LSTM decoder, and selected the MLP decoder for position decoding models, while selecting the LSTM decoder for velocity decoding models. Second, the reported Tracking accuracy of the position decoding model was quite low: we find that a model which simply predicts the ground-truth initial position at all timesteps achieves comparable performance (Table \ref{tab:tracking_results}).

To mitigate training instabilities without requiring an extensive hyperparameter search, we adopt a more modern default optimization regime (i.e.\ Adam without weight decay $\rightarrow$ AdamW with default weight decay; fixed learning rate $\rightarrow$ the learning-rate schedule from \citet{sivakumar-2024}). We then train 9 seeds of a position decoding model using the exact architecture from \citet{salter-2024}, but with the LSTM decoder. While several seeds do converge, training remains unstable, with many seeds becoming trapped in low-movement local minima in which the predicted trajectories exhibit near-zero movement (Figure \ref{fig:fig1}, red traces in the left column). These minima correspond to a trivial solution in which the model predicts near the ground-truth initial position throughout the sample window, neglecting the sEMG signals that would otherwise inform subsequent movement. To alleviate this failure mode, we tune a single previously neglected hyperparameter: a scalar applied to the decoder outputs. \citet{salter-2024} fixed this scalar at 0.01. Simply increasing it to 0.1 yields stable training that bypasses these low-movement minima (Figure \ref{fig:fig1}, green traces in the left column).

On the Regression task, we observe a related but distinct failure mode when using the default output scalar of 0.01 for position decoding models. Again, the models often become trapped in low-movement minima; however, because the Regression task does not provide a ground-truth initial position with which to seed trajectories, the models instead collapse to an even more trivial solution: predicting near the per-joint median angle with virtually no movement. Here too, increasing the output scalar restores stable training. In this case, $s=1$ was required, as $s=0.1$ still exhibited instabilities.

These observations help explain why position decoding underperformed in the original emg2pose experiments of \citet{salter-2024}: under a fixed output scalar that frequently induces optimization collapse in LSTM-based position models, the hyperparameter search instead favored an MLP decoder. In all subsequent experiments, we train all models with the LSTM decoder, AdamW with weight decay $10^{-2}$, the learning-rate schedule from \citet{sivakumar-2024}, and an output-scalar sweep over $\{0.01, 0.1, 1\}$, selecting the best value by validation loss and verifying stable convergence by running eight additional seeds and confirming that none exhibit instabilities. When reporting results, we take the average performance across seeds per each user, and then take the mean and standard deviation across users.

\subsection{Position decoding outperforms velocity decoding for the Tracking task}

Having identified a stable training regime for position decoding models, we next ask whether they still underperform velocity decoding models on the Tracking task. Somewhat surprisingly, we find the opposite: position decoding models significantly outperform velocity decoding models on the Tracking task across all generalization conditions and error metrics. Notably, both our position and velocity decoding models improve upon the corresponding results reported by \citet{salter-2024}, indicating that our updated optimization regime benefits both model classes rather than uniquely favoring position decoding models.

To contextualize these improvements, we also include a "Static" baseline, which simply predicts the ground-truth initial position throughout each sample window. Relative to this baseline, our position decoding models eliminate substantially more error than our velocity decoding models and prior baselines. At the same time, comparison to the Static baseline suggests that even the best-performing models remove only a modest fraction of the total error on this dataset, highlighting that the Tracking task remains far from solved.

We also investigated whether multi-task training on a combination of Tracking and Regression improves performance on the Tracking task itself, but generally did not observe a benefit. This is unsurprising: the main capability that Regression training might contribute is the ability to infer the next position without a reliable estimate of the previous one, whereas the Tracking task already provides an anchored initial position. As a result, the additional supervision from Regression appears to offer little advantage for Tracking.

\begin{table}[h!]
\centering
\caption{Tracking task results (mean $\pm$ sd across users).}
\label{tab:tracking_results}
\begin{tabular}{lcccccc}
\toprule
& \multicolumn{2}{c}{User-Stage} & \multicolumn{2}{c}{User} & \multicolumn{2}{c}{Stage} \\
\cmidrule(lr){2-3} \cmidrule(lr){4-5} \cmidrule(lr){6-7}
Model & AE ($^\circ$) & LD (mm) & AE ($^\circ$) & LD (mm) & AE ($^\circ$) & LD (mm) \\
\midrule
Static & $14.29{\scriptscriptstyle\,\pm\,1.86}$ & $21.48{\scriptscriptstyle\,\pm\,2.33}$ & $10.48{\scriptscriptstyle\,\pm\,1.81}$ & $15.71{\scriptscriptstyle\,\pm\,2.98}$ & $14.86{\scriptscriptstyle\,\pm\,1.91}$ & $21.81{\scriptscriptstyle\,\pm\,2.58}$ \\
\cmidrule(lr){1-7}
\multicolumn{7}{l}{\textbf{Salter et al.}} \\
\addlinespace[0.2em]
emg2pose & $15.05{\scriptscriptstyle\,\pm\,1.22}$ & $20.69{\scriptscriptstyle\,\pm\,1.39}$ & $11.33{\scriptscriptstyle\,\pm\,1.10}$ & $14.75{\scriptscriptstyle\,\pm\,1.53}$ & $15.39{\scriptscriptstyle\,\pm\,1.50}$ & $20.88{\scriptscriptstyle\,\pm\,1.92}$ \\
vemg2pose & $11.03{\scriptscriptstyle\,\pm\,1.01}$ & $15.35{\scriptscriptstyle\,\pm\,1.46}$ & $7.70{\scriptscriptstyle\,\pm\,0.98}$ & $10.22{\scriptscriptstyle\,\pm\,1.54}$ & $11.20{\scriptscriptstyle\,\pm\,1.43}$ & $15.18{\scriptscriptstyle\,\pm\,1.81}$ \\
\cmidrule(lr){1-7}
\multicolumn{7}{l}{\textbf{Ours}} \\
\addlinespace[0.2em]
Pos ST & $10.25{\scriptscriptstyle\,\pm\,0.93}$ & $14.18{\scriptscriptstyle\,\pm\,1.35}$ & $7.35{\scriptscriptstyle\,\pm\,0.92}$ & $9.67{\scriptscriptstyle\,\pm\,1.48}$ & $10.21{\scriptscriptstyle\,\pm\,1.32}$ & $13.73{\scriptscriptstyle\,\pm\,1.69}$ \\
Vel ST& $10.75{\scriptscriptstyle\,\pm\,0.93}$ & $14.82{\scriptscriptstyle\,\pm\,1.40}$ & $7.64{\scriptscriptstyle\,\pm\,0.94}$ & $10.02{\scriptscriptstyle\,\pm\,1.52}$ & $10.76{\scriptscriptstyle\,\pm\,1.34}$ & $14.43{\scriptscriptstyle\,\pm\,1.71}$ \\
Pos MT & $10.29{\scriptscriptstyle\,\pm\,0.97}$ & $14.22{\scriptscriptstyle\,\pm\,1.37}$ & $7.36{\scriptscriptstyle\,\pm\,0.90}$ & $9.68{\scriptscriptstyle\,\pm\,1.46}$ & $10.28{\scriptscriptstyle\,\pm\,1.33}$ & $13.81{\scriptscriptstyle\,\pm\,1.70}$ \\
Vel MT & $10.75{\scriptscriptstyle\,\pm\,0.97}$ & $14.81{\scriptscriptstyle\,\pm\,1.41}$ & $7.64{\scriptscriptstyle\,\pm\,0.95}$ & $10.00{\scriptscriptstyle\,\pm\,1.53}$ & $10.77{\scriptscriptstyle\,\pm\,1.36}$ & $14.44{\scriptscriptstyle\,\pm\,1.75}$ \\
\bottomrule
\end{tabular}%

\end{table}

\subsection{Multi-task training improves performance on the Regression task for both position and velocity decoding models}

On the Regression task, we observe smaller performance differences between position and velocity decoding models than on Tracking. Under single-task training, position decoding models slightly underperform velocity decoding models. However, the dominant effect on this task is not output parameterization, but training regime: switching from single-task to multi-task training yields a substantial improvement for both model classes, at which point position decoding models perform on par with velocity decoding models.

As in the Tracking task, we compare against a "Static" baseline. The Static baseline for the Regression task, which merely predicts the training-set median angle of each joint at all timesteps, helps put our gains into perspective: though our multi-task models achieve substantial improvements relative to our single-task models and prior baseline, even the best models still leave a substantial fraction of the total error unexplained.

The strong benefit of multi-task training suggests that supplementing Regression with the Tracking task provides a useful regularizing effect. Unlike Regression, Tracking supplies an anchored initial position, which may make the underlying hand-motion dynamics easier to learn. Joint training may therefore help the model acquire a more stable representation of these dynamics, improving performance on the less constrained Regression task.

\begin{table}[h!]
\centering
\caption{Regression task results (mean $\pm$ sd across users).}
\label{tab:regression_results}
\begin{tabular}{lcccccc}
\toprule
& \multicolumn{2}{c}{User-Stage} & \multicolumn{2}{c}{User} & \multicolumn{2}{c}{Stage} \\
\cmidrule(lr){2-3} \cmidrule(lr){4-5} \cmidrule(lr){6-7}
Model & AE ($^\circ$) & LD (mm) & AE ($^\circ$) & LD (mm) & AE ($^\circ$) & LD (mm) \\
\midrule
Static & $18.87{\scriptscriptstyle\,\pm\,2.01}$ & $28.77{\scriptscriptstyle\,\pm\,2.78}$ & $16.86{\scriptscriptstyle\,\pm\,1.80}$ & $25.17{\scriptscriptstyle\,\pm\,2.84}$ & $19.18{\scriptscriptstyle\,\pm\,1.71}$ & $28.57{\scriptscriptstyle\,\pm\,2.46}$ \\
\cmidrule(lr){1-7}
\multicolumn{7}{l}{\textbf{Salter et al.}} \\
\addlinespace[0.2em]
emg2pose & $15.57{\scriptscriptstyle\,\pm\,1.30}$ & $21.49{\scriptscriptstyle\,\pm\,1.71}$ & $12.57{\scriptscriptstyle\,\pm\,1.30}$ & $16.28{\scriptscriptstyle\,\pm\,1.81}$ & $15.17{\scriptscriptstyle\,\pm\,1.59}$ & $20.53{\scriptscriptstyle\,\pm\,2.13}$ \\
vemg2pose & $15.63{\scriptscriptstyle\,\pm\,1.37}$ & $21.30{\scriptscriptstyle\,\pm\,1.87}$ & $12.24{\scriptscriptstyle\,\pm\,1.34}$ & $15.82{\scriptscriptstyle\,\pm\,1.90}$ & $15.22{\scriptscriptstyle\,\pm\,1.58}$ & $20.38{\scriptscriptstyle\,\pm\,2.07}$ \\
\cmidrule(lr){1-7}
\multicolumn{7}{l}{\textbf{Ours}} \\
\addlinespace[0.2em]
Pos ST & $15.47{\scriptscriptstyle\,\pm\,1.30}$ & $21.29{\scriptscriptstyle\,\pm\,1.82}$ & $12.19{\scriptscriptstyle\,\pm\,1.27}$ & $15.79{\scriptscriptstyle\,\pm\,1.80}$ & $14.96{\scriptscriptstyle\,\pm\,1.63}$ & $20.23{\scriptscriptstyle\,\pm\,2.15}$ \\
Vel ST & $15.35{\scriptscriptstyle\,\pm\,1.32}$ & $20.92{\scriptscriptstyle\,\pm\,1.87}$ & $12.17{\scriptscriptstyle\,\pm\,1.31}$ & $15.58{\scriptscriptstyle\,\pm\,1.85}$ & $14.62{\scriptscriptstyle\,\pm\,1.65}$ & $19.62{\scriptscriptstyle\,\pm\,2.21}$ \\
Pos MT & $14.58{\scriptscriptstyle\,\pm\,1.29}$ & $19.78{\scriptscriptstyle\,\pm\,1.83}$ & $11.54{\scriptscriptstyle\,\pm\,1.19}$ & $14.74{\scriptscriptstyle\,\pm\,1.64}$ & $14.02{\scriptscriptstyle\,\pm\,1.62}$ & $18.61{\scriptscriptstyle\,\pm\,2.16}$ \\
Vel MT & $14.63{\scriptscriptstyle\,\pm\,1.34}$ & $19.77{\scriptscriptstyle\,\pm\,1.95}$ & $11.64{\scriptscriptstyle\,\pm\,1.25}$ & $14.74{\scriptscriptstyle\,\pm\,1.74}$ & $13.85{\scriptscriptstyle\,\pm\,1.62}$ & $18.33{\scriptscriptstyle\,\pm\,2.16}$ \\
\bottomrule
\end{tabular}%
\end{table}

\subsection{Position decoding models exhibit greater local jitter, while multi-task training improves temporal coherence}
\label{sec:freq}
 
\begin{figure}[h!]
  \centering
  \includegraphics[height=6cm]{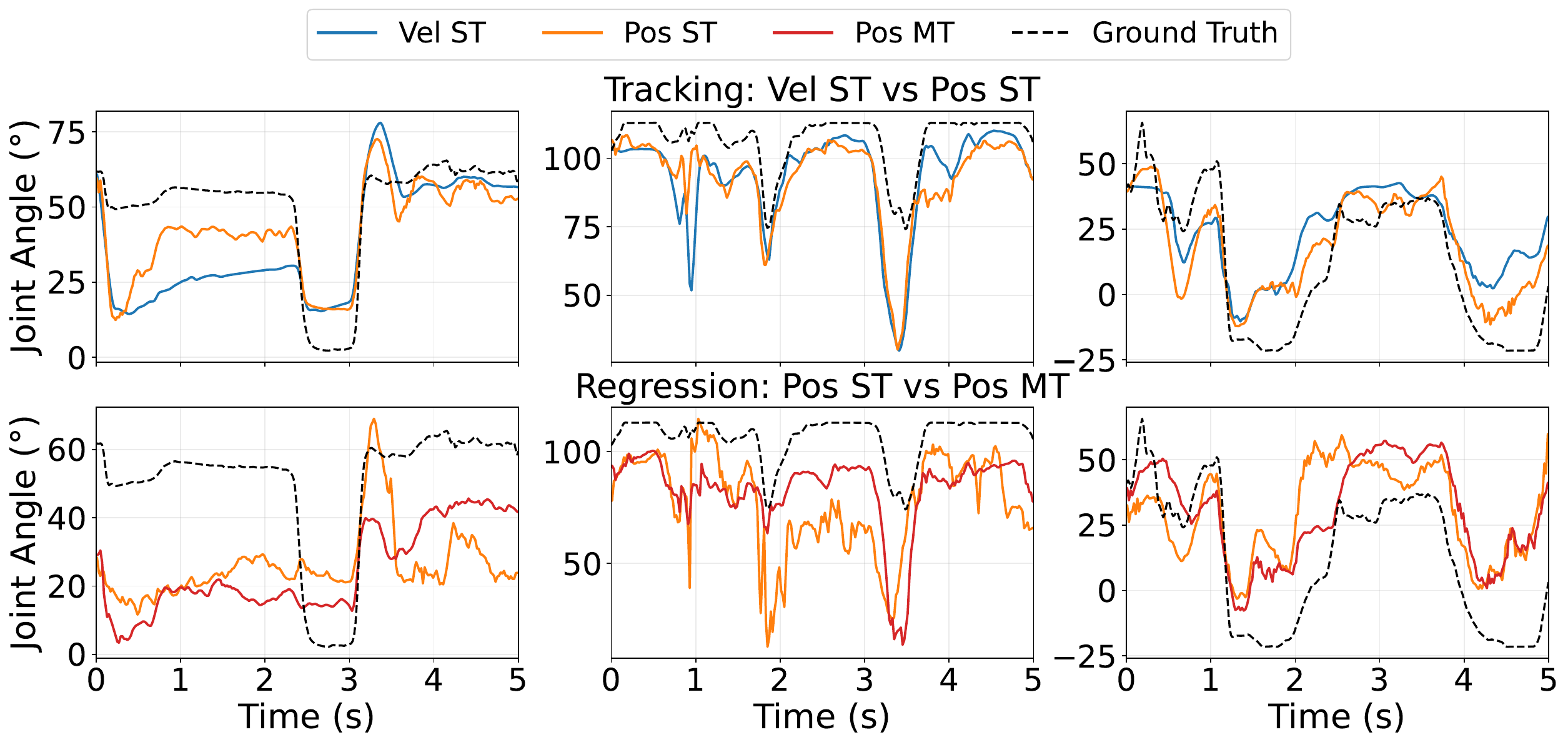}
  
  \caption{\textbf{Representative trajectory examples illustrate qualitative differences between model classes.}
Top row: on the Tracking task, position decoding models often appear more locally jagged than velocity decoding models. Bottom row: on the Regression task, multi-task position decoding models exhibit more coherent local dynamics than single-task position decoding models. Dashed black lines show ground-truth trajectories. Figure \ref{fig:fig2} quantifies these effects across sample windows and frequency bands.}
  \label{fig:fig2_prelude}
\end{figure}

To build intuition for the effects of output parameterization and multi-task training, we first show representative predicted trajectories (Figure \ref{fig:fig2_prelude}). On the Tracking task, position decoding models often appear more locally jagged than velocity decoding models. On the Regression task, multi-task models exhibit more coherent local dynamics than their single-task counterparts. We next quantify these qualitative differences across all sample windows using timestep-wise and frequency-domain analyses.

\begin{figure}[h!]
  \centering
  \includegraphics[height=12cm]{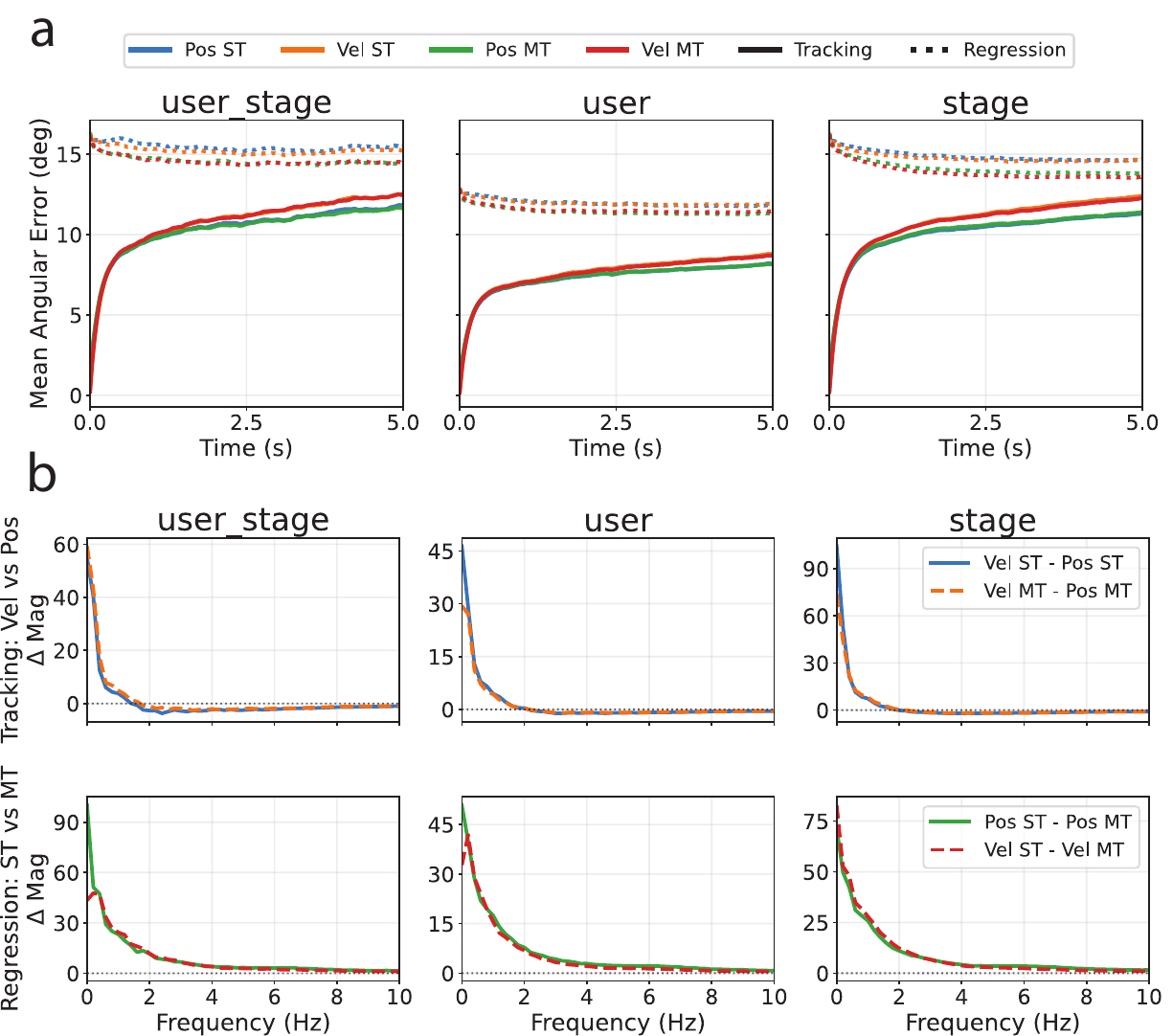}
  
  \caption{\textbf{Error dynamics across time and frequency reveal complementary effects of output parameterization and multi-task training.} 
\textbf{(a)} Mean angular error as a function of time within sample windows for Tracking (solid) and Regression (dotted) under the user\_stage, user, and stage generalization conditions. Tracking models begin from near-zero error and accumulate error over time, with position decoding models generally accumulating error more slowly than velocity decoding models. In contrast, Regression models generally improve over time as additional sEMG context becomes available, with multi-task models showing faster error reduction than single-task models. 
\textbf{(b)} Frequency-domain comparison of model residuals. For each model, we compute residuals relative to the ground-truth targets, take the absolute value of the FFT coefficients of those residuals, and then plot pairwise differences between models. Positive values indicate lower residual magnitude for the second model named in the comparison, while negative values indicate lower residual magnitude for the first. Top row: comparing velocity and position decoding models on Tracking shows that position decoding reduces low-frequency residuals but increases high-frequency residuals, consistent with improved robustness to drift but greater high-frequency jitter. Bottom row: comparing single-task and multi-task models on Regression shows that multi-task training reduces residuals broadly across frequencies.}
  \label{fig:fig2}
\end{figure}

As expected, Tracking models begin from near-zero error at the first timesteps of sample windows, but quickly accumulate error over the course of the window (Figure \ref{fig:fig2}a, solid lines). We observe that position decoding models generally accumulate error more slowly than velocity decoding models across all generalization conditions. This is consistent with position decoding being more robust to drift, which is intuitive: unlike velocity decoding models, position decoding models do not need to continually adjust their output distribution at each timestep to compensate for past errors.

For the Regression task, all models generally reduce error over the course of the sample windows (Figure \ref{fig:fig2}a, dashed lines). This is unsurprising, since there is no ground-truth initial position to drift away from in this task, and accumulating sEMG context can support more informed predictions of hand position. We observe that multi-task models generally reduce error more rapidly than single-task models, especially early in the sample windows. This likely stems from the Tracking task providing a stable reference point that helps the model learn hand motion dynamics more effectively. Multi-task training may therefore act as a form of curriculum, in which the more anchored and comparatively easier Tracking task regularizes performance on the more challenging Regression task.

Moving beyond overall accuracy, we next ask which components of the targets are better captured by our models on the two tasks. For these analyses, we first compute residuals between each model's predictions and the ground-truth targets, then perform a Fast Fourier Transform on those residuals and take the absolute value of the resulting coefficients. This gives an indication of the magnitude of model error across frequency bands. Finally, when comparing two models, we compute the difference between their absolute FFT coefficients.

We begin by comparing velocity decoding models against position decoding models (Figure \ref{fig:fig2}b, top row). Consistent with the qualitative examples in Figure \ref{fig:fig2_prelude}, position decoding models generally have lower-magnitude residuals than velocity decoding models at lower frequencies (up to \textasciitilde1.5 Hz), consistent with better capture of slower, intentional hand movements. However, position decoding models also exhibit higher-magnitude residuals at higher frequencies, consistent with greater high-frequency jitter in their predictions.

We then compare multi-task models against single-task models on the Regression task. Again consistent with the qualitative examples in Figure \ref{fig:fig2_prelude}, multi-task training appears to provide a broad benefit by reducing residuals across all frequencies, suggesting that the learned dynamical priors improve the model's underlying mapping from sEMG to pose rather than merely correcting specific low-frequency drift or high-frequency jitter.

\subsection{A computationally trivial speed-adaptive causal filter effectively bypasses the smoothness--accuracy tradeoff}

Having found that position decoding models yield more accurate but also more jittery predictions, we next ask whether a computationally trivial causal filter can bypass this tradeoff. Standard causal low-pass filters can introduce considerable lag. Inspired by the One Euro filter \citep{casiez20121}, which uses a speed-adaptive cutoff to suppress jitter while preserving fast movements, we instead use an even simpler variant (Methods \ref{filter_method}) with only a single parameter, $\beta$, that controls the tradeoff between smoothness and lag. We sweep a range of $\beta$ values for all models and, for each generalization condition, plot accuracy against the filtered model's mean speed. We focus on mean speed because it was the smoothness metric used by \citet{salter-2024}. More generally, among models with similar accuracy, lower-speed trajectories may be preferable for embodied applications if they reflect less unnecessary movement.

\begin{figure}[t!]
  \centering
  \includegraphics[height=12cm]{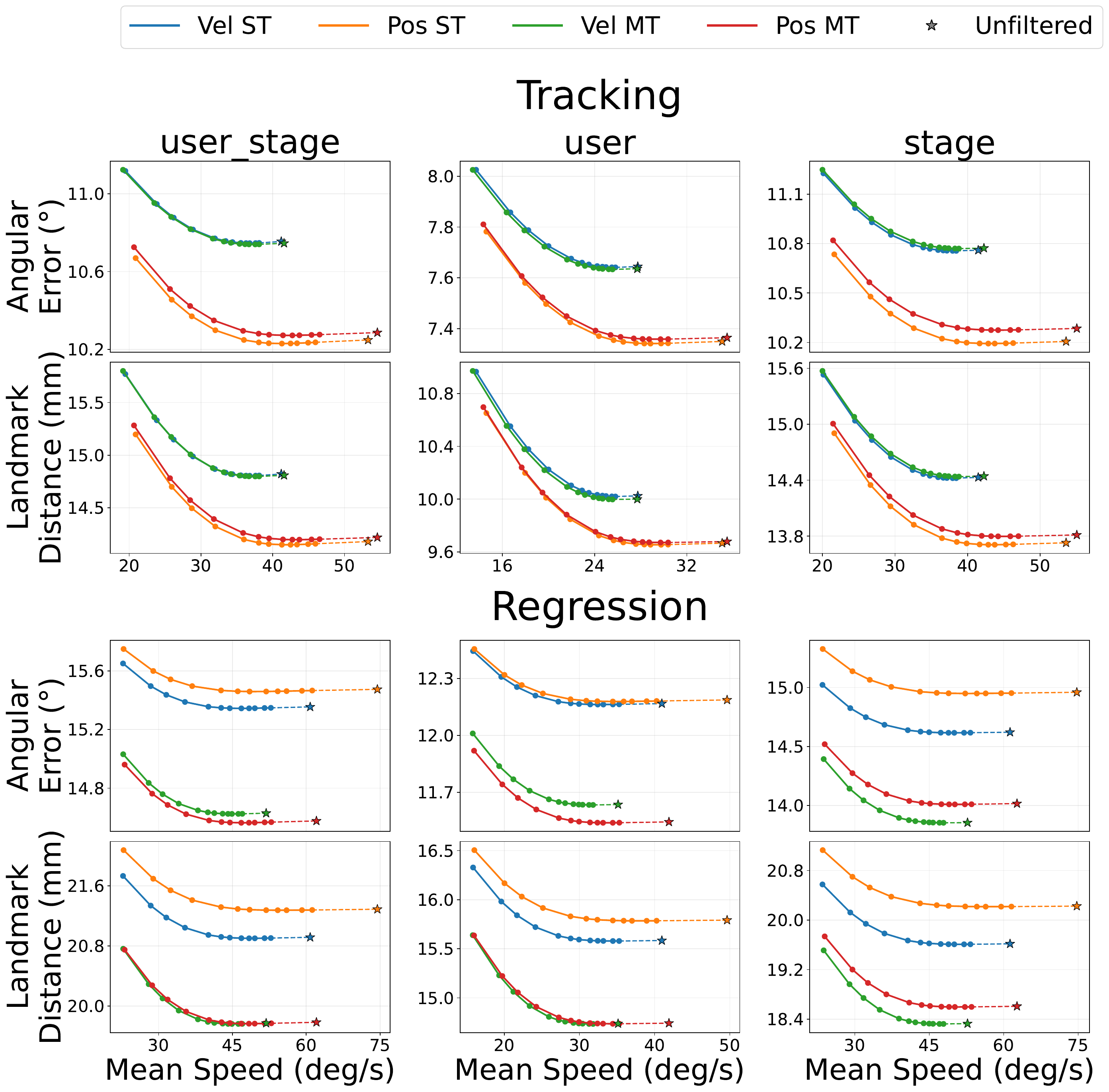}
  
  \caption{\textbf{A simple speed-adaptive causal filter improves smoothness while preserving accuracy.}
For each model, we sweep the filter parameter $\beta$ and plot accuracy against the mean speed of the resulting filtered trajectories under the user\_stage, user, and stage generalization conditions. Star markers denote the unfiltered models. Top two rows: Tracking task results for angular error and landmark distance. Although unfiltered position decoding models have higher mean speed than velocity decoding models, filtering allows them to achieve lower error across the full observed range of mean speeds, effectively bypassing the apparent smoothness--accuracy tradeoff. Bottom two rows: Regression task results. Multi-task training substantially lowers mean speed relative to single-task training for both output parameterizations, and filtering further reduces speed while largely preserving accuracy. At matched smoothness levels, multi-task models consistently outperform single-task models, while position and velocity decoding models show near-parity.}
  \label{fig:fig3}
\end{figure}

We begin with the Tracking task (Figure \ref{fig:fig3}, top two rows). Without filtering, position decoding models indeed produce trajectories with higher mean speed than velocity decoding models, consistent with less smooth predictions and greater high-frequency jitter. However, after filtering both model classes across a range of $\beta$ values, position decoding models remain more accurate than velocity decoding models across the full observed range of mean speeds and for both error metrics. In this sense, the smoothness--accuracy tradeoff can be effectively bypassed: rather than accepting a loss in accuracy to obtain smoother trajectories, one can instead filter the more accurate position decoding models and still outperform velocity decoding models at matched or lower mean speeds. These results suggest that velocity decoding may be a suboptimal choice for obtaining trajectories that are both smooth and accurate on the Tracking task.

We then turn to the Regression task (Figure \ref{fig:fig3}, bottom two rows). Here, multi-task training itself substantially reduces mean speed for both position and velocity decoding models, consistent with the more coherent local dynamics observed in Section \ref{sec:freq}. Moreover, for all models, filtering further reduces mean speed while largely preserving accuracy. At comparable levels of smoothness, position and velocity decoding models show near-parity in accuracy, whereas multi-task models consistently outperform their single-task counterparts. Thus, on the Regression task, the main practical takeaway is that smooth predictions can be obtained for all model classes, but multi-task training yields a more favorable smoothness--accuracy profile regardless of output parameterization.

\section{Discussion}

Our results revisit a central conclusion of the original emg2pose benchmark study \citep{salter-2024}: that velocity decoding is preferable to position decoding for real-time hand-pose prediction from sEMG. We find that this conclusion is not robust. Once position decoding models are trained in a stable regime, they substantially outperform velocity decoding models on the Tracking task across all generalization conditions and both error metrics. At the same time, this advantage is not unconditional: position decoding models also exhibit greater high-frequency jitter, suggesting a tradeoff between drift robustness and local smoothness. Importantly, however, we find that this tradeoff can be effectively bypassed in practice using a computationally trivial speed-adaptive causal filter, which allows filtered position decoding models to outperform velocity decoding models across the full observed range of mean speeds on the Tracking task. Together, these findings suggest that direct position decoding, rather than velocity decoding, may be the stronger default choice for this benchmark.

A key contribution of this work is to show that apparently high-level modeling conclusions can hinge on low-level optimization details. In particular, we find that LSTM-based position decoding models are highly sensitive to the scale of decoder outputs, and that the default scalar used in \citet{salter-2024} frequently leads to optimization collapse into low-movement solutions. On the Tracking task, this collapse manifests as trajectories that remain close to the ground-truth initial position; on the Regression task, it manifests as trajectories that remain near per-joint median angles with little movement. Simply tuning this neglected scalar is sufficient to recover stable training. More broadly, these results reinforce a familiar but important lesson for benchmark design: conclusions about modeling strategies should be drawn only after ensuring that the compared systems have each been placed in a reasonably competent training regime. Otherwise, optimization failures can masquerade as principled architectural advantages.

Our analyses also clarify that the choice between position and velocity decoding is not just a matter of overall error, but of the kinds of errors models make. On the Tracking task, position decoding models accumulate error more slowly over time and reduce low-frequency residuals relative to velocity decoding models, consistent with greater robustness to drift. At the same time, they exhibit larger residuals at higher frequencies, consistent with greater local jitter. This distinction is important for embodied applications. A decoder that is more accurate in aggregate but introduces high-frequency wobble may still require downstream smoothing, while a decoder that is smoother but drifts more may be harder to stabilize over longer horizons. Our filtering results suggest that, at least for this benchmark, the former regime is preferable: high-frequency jitter is comparatively easy to suppress with lightweight post-processing, whereas the drift associated with velocity decoding is harder to recover from.

On the Regression task, the picture is somewhat different. Here, differences between position and velocity decoding models are modest, especially once multi-task training is introduced. The strongest effect is instead the benefit of jointly training on Tracking and Regression, which substantially improves Regression performance and yields more temporally coherent predictions. One interpretation is that Tracking provides an easier, more anchored learning problem that helps the model acquire better dynamical priors for hand motion, which then transfer to the less constrained Regression setting. In this sense, multi-task training may function as a simple form of curriculum. At the same time, the velocity decoding model in the Regression setting should not be interpreted as a comparison against a "pure" velocity prediction strategy: because a position decoding head is used for first 250 ms, the model is better viewed as a hybrid output parameterization rather than as a strict velocity decoding model. This nuance may help explain why the Regression task yields a less dramatic position-versus-velocity contrast than the Tracking task.

We also note some inconsistencies in the relative ordering of models across generalization conditions on the Regression task, particularly when comparing the user\_stage, user, and stage splits. We place greater weight on the user\_stage and user conditions, since generalization across users is more central to practical deployment than generalization across held-out recording stages alone. The stage condition remains useful, but it is arguably the least consequential of the three for assessing whether a model can generalize to new people. In this context, the most important conclusion is not the precise ranking of position and velocity decoding models on every split, but the broader pattern that multi-task training consistently improves Regression performance and that output parameterization matters less here than on Tracking.

Several limitations should be kept in mind. First, our study remains centered on the architectures and benchmark formulation introduced by \citet{salter-2024}. While this focus is valuable for revisiting their conclusions under a better-controlled optimization regime, it does not exhaust the space of possible decoders. In particular, there may be alternative ways of encouraging smoothness directly during training, such as explicit temporal smoothness penalties, frequency-domain losses, uncertainty-aware objectives, or architectures designed to better separate low- and high-frequency components of motion. Second, although our speed-adaptive filter is attractive precisely because of its negligible computational cost, it is still a post hoc solution rather than a property learned by the decoder itself. Future work could explore whether similarly favorable smoothness--accuracy frontiers can be achieved end-to-end.

Our results also point toward richer forms of curriculum learning. Here, we considered only a simple joint objective over Tracking and Regression. But the observed benefits of multi-task training suggest that more structured training schedules may be fruitful: for example, pretraining on anchored Tracking and then fine-tuning on Regression, gradually reducing access to reliable initial pose information, or mixing tasks in a stage-dependent way. Such strategies may prove especially useful for settings in which the target task is weakly constrained or where robust local dynamics matter more than instantaneous pointwise accuracy.

Finally, these findings have implications beyond this specific benchmark. In fast-moving applied machine learning domains, it is tempting to draw broad conclusions from benchmark leaderboards about which parameterizations, architectures, or output spaces are intrinsically best. Our results caution that such conclusions may be surprisingly brittle when the optimization regime is not equally well matched to all compared models. In the present case, a neglected scaling hyperparameter and a trivial causal filter were sufficient to overturn the benchmark's main modeling conclusion on its flagship Tracking task. We therefore view this work not only as a revision of the emg2pose baseline, but also as a reminder that careful optimization and mechanistic error analysis remain essential for drawing reliable lessons from benchmark results.

\bibliographystyle{plainnat}
\bibliography{references}

\end{document}